\documentclass[conference]{IEEEtran}
\ifCLASSINFOpdf
\else
\fi
\IEEEoverridecommandlockouts
\usepackage{cite}
\usepackage{amsmath,amssymb,amsfonts}
\usepackage{algorithmic}
\usepackage{graphicx}
\usepackage{textcomp}
\usepackage{xcolor}

\usepackage{times}
\usepackage{graphicx}
\usepackage{booktabs} 
\usepackage{graphicx}
\usepackage{epstopdf}
\usepackage{amsmath}
\usepackage{graphicx}
\usepackage{multicol}
\usepackage{enumitem}
\usepackage{comment}
\usepackage{array}
\usepackage{multirow} 
\usepackage{epstopdf}
\usepackage{color}
\usepackage{diagbox}
\usepackage{latexsym} 
\usepackage{times}
\usepackage{graphicx} 
\usepackage{float} 
\usepackage{mathtools, amsfonts, amssymb, amsthm,color,bm} 
\usepackage{cite}
\usepackage[english]{babel}
\usepackage[figurename=Fig.,font={small,it}]{caption} 
\usepackage{siunitx}
\usepackage{algorithm}
\usepackage{algorithmic}  
\usepackage{graphicx}

\def\S{{\mathcal{S}}}
\def\N{{\mathcal{N}}}
\def\E{{\mathcal{E}}}

\usepackage[numbers]{natbib} 
\begin{document}
\title{Physics-Informed  Learning for High Impedance Faults Detection  \thanks{The authors acknowledge the support from the Department of Energy through the 
		Advanced Grid Modeling (AGM) Program, and the Center for Non Linear Studies (CNLS) at Los Alamos National Laboratory.} }

\author{\IEEEauthorblockN{Wenting~Li}
	\IEEEauthorblockA{\textit{ Center for Non-Linear Studies (CNLS)} \\
		\textit{Los Alamos National Laboratory}\\
		Los Alamos, NM \\
		wenting@lanl.gov}
	\and
	\IEEEauthorblockN{Deepjyoti~Deka}
	\IEEEauthorblockA{\textit{Theoretical Division  } \\
		\textit{Los Alamos National Laboratory}\\
		Los Alamos, NM \\
		deepjyoti@lanl.gov} 
}
  
\maketitle

\begin{abstract}
	High impedance faults (HIFs) in distribution grids may cause wildfires and threaten human lives. Conventional protection relays at substations fail to detect more than 10\% HIFs since over-currents are low and the signatures of HIFs are local. With more $\mu$PMU being installed in the distribution system,  high-resolution $\mu$PMU datasets provide the opportunity of detecting HIFs from multiple points. Still, the main obstacle in applying the $\mu$PMU datasets is the lack of labels. To address this issue, we construct a  physics-informed convolutional auto-encoder (PICAE) to detect HIFs without labeled HIFs for training. The significance of our PICAE is a physical regularization, derived from the elliptical trajectory of voltages-current characteristics, to distinguish HIFs from other abnormal events even in highly noisy situations. We formulate a system-wide detection framework that merges multiple nodes' local detection results to improve the detection accuracy and reliability. The proposed approaches are validated in the IEEE 34-node test feeder simulated through PSCAD/EMTDC. Our PICAE outperforms the existing works in various scenarios and is robust to different observability and noise.
\end{abstract}

 
\IEEEpeerreviewmaketitle

\section{Introduction} 
Energized conductors hitting the high impedance ground surfaces, usually accompanied by arc flashing, have led to most HIFs \cite{GB12}. People are concerned with HIFs, as they are one of the main causes/initiators of destructive wildfires and threaten public safety.  Diversity of physical models have well described the process of HIFs of randomness and nonlinearity \cite{MRB96}. However, more than 10\% detection failures of HIFs have been reported \cite{GMGS14} using voltages or currents measured by devices at relays or breakers \cite{AWTJ06}.   Conventional over-current protection systems often neglect HIFs due to the low fault current \cite{GB12, GMGS14}. This problem is acerbated in distribution grids as measurements are not ubiquitous, and signatures of HIFs are local and do not propagate much in the grid. In recent years, there has been growing interest in detecting HIFs in distribution grids accurately when more $\mu$PMUs being installed.

The existing data-driven HIF detection methods usually separate HIFs from others by supervised classification with various features, based on time-domain, frequency-domain, and time-frequency domain measurements \cite{GB12, WGD16,  CD18, GMGS14}. However, these methods are either not robust to noise and low harmonics rates or require a sufficient number of labeled datasets\footnote{Labeled datasets denote the types of the recorded datasets are tagged. }  to learn the features.  

To address these issues, we propose a novel and practical HIF neural-network-based detector for distribution grids with limited measurement availability that uses only normal data \emph{and no labeled faults} for training. Neural networks have achieved great success in computer vision, natural language processing, and health care \cite{LBHG15}. While applications with labeled data are many, success with partially labeled or even completely unlabeled datasets has been demonstrated with satisfactory accuracy and efficiency \cite{PGHY16, MSY16}. One label-free model is Auto-encoder (AE) \cite{LBHG15}, a neural network architecture consisting of an encoder and a decoder to learn the features and reconstruct the data.   Various AE derivatives have been proposed for specific applications \cite{PGHY16, MSY16}. However, such pure data-driven applications take the risk of violating the physical rules that govern the cyber-physical systems such as power grids.  Hence, our method can overcome these issues by judiciously using constraints related to the \emph{physics-informed dynamics} in regular operation during the detector training.  

We are inspired by the recent attempts of embedding physical laws in neural networks or statistical machine learning for power flow calculation, state estimation, topology learning \cite{ZS20, PDC18, TDDMCS20}, and power system monitoring \cite{LWC18, LDCW19}. Outside of power grids, \cite{RPK17} reveals promising progress in regulating the learned parameters of neural networks with physical laws as priors. These physics-based promotions improve both interpretability as well as the model's computational efficiency.  

\textbf{Contribution:} 
We propose a physics-informed learning framework to detect HIFs, on the conditions of a limited number of measured nodes and scarce labeled faults for training. Explicitly, relying on the fact that elliptical curves can model the trajectory of normal voltage-current with time, we construct a Convolutional Auto-Encoder (CAE) to represent the voltage time-series data during normal operations (no faults). Additionally, we constrain its output with the physics-regulated (PR) elliptical characteristics of voltages and currents. Furthermore, as HIF's signatures are local,  we establish a low-communication central scheme that merges the observed nodes' local decisions to augment the detection robustness and reliability. We validate the proposed methods in the IEEE 34 node benchmark system \cite{MS07} simulated by  {Power Systems Computer-Aided Design (PSCAD) \cite{PSCAD}}. We demonstrate our detector's high performance even when systems are not fully observed and interpret the physics-informed regularization's advantages to distinguish HIFs from others. Moreover, we show that PICAE outperforms existing schemes on HIF detection in multiple noisy scenarios.  

The remaining part of this paper is organized as follows: Section \ref{sec:model} introduces the physical rules of HIFs; based on these rules, we construct a physics-informed convolutional autoencoder (PICAE) to detect HIFs in Section \ref{sec:method}; the detection framework of local and central determination are presented in Section \ref{sec:detect};  numerical experiments in Section \ref{sec:simu} show the detection performance of the proposed approaches, in comparison with some existing works in different scenarios. Section \ref{sec:con} concludes the main results.

\section{Background of the Physical Model for HIF}\label{sec:model}
HIF is a nonlinear, random event that is often unnoticeable by over-current relays or fuses. In the last decades, various arc models have been utilized to describe the stable or dynamic HIF process \cite{SS90, MRB96, GB12}. Two-parallel diodes and a voltage source model accurately represent the dynamic re-striking and quenching process of arcs during HIF \emph{at the fault point}. \cite{GB12,WGD16}.  

\subsection{Modeling of HIF Process} 
Let $v(t)$ be the single phase voltage at the time $t$ that interacts with the two DC voltage sources $V_p >0, V_n <0 $, and variable resistances $R_p \neq R_n$ in the down and up lines. 
\begin{align}\label{v_HIF}
	v(t) = \begin{cases}
		V_p + i_p(t)R_p & \quad \text{if } v(t) > V_p \\
		V_n - i_n(t)R_n & \quad \text{if } v(t)< V_n \\
		v(t-1) & \quad \text{else}
	\end{cases} 
\end{align} 
When $v(t) > V_p$, the diode $D_p$ is switch on to allow fault current $i_p$ to flow through, and when $v(t)< V_n$, the diode $D_n$ is switch on to let $i_n$ flow in. These structures mimic the re-striking process of arcs; otherwise, no currents flow through the HIF circuit and the voltages of the fault point keep the same with the previous phase voltage $v(t-1)$, which represents the quenching of arcs. Note that the re-striking and quenching process will \emph{cycle and last} for seconds or even longer \cite{TPV18}. This process is random and nonlinear since the impedance $R_n, R_p$ are randomly varying. 

\subsection{Physical Laws of HIFs} \label{physical} 
On normal conditions, it is demonstrated that the trajectories of voltages and currents are rotated ellipses for resistance-inductive or resistance-capacitive linear circuits, and are circles if resistance is zero \cite{WGD16}. Let phase voltages and currents be $v(t) = V_0\cos (\omega t), c(t) = C_0 \cos (\omega t - \phi)$ with a phase angle $\phi$, then we can fit them into the standard parametric format of a rotated ellipse equation as follows:
\begin{equation}\label{parametric}
	(\dfrac{v(t)}{\alpha_1 } + \dfrac{c(t)}{\alpha_2 } )^2+ (\dfrac{v(t)}{\alpha_3 } - \dfrac{c(t)}{\alpha_4 })^2 = 1
\end{equation}
where $\alpha_1 = 2V_0 \cos ( {\phi}/{2}), \alpha_2 = 2C_0 \cos ( {\phi}/{2}), \alpha_3 = 2V_0 \sin ( {\phi}/{2}), \alpha_4 = 2C_0 \sin ( {\phi}/{2}) $, where $\alpha_i$ are determined by line impedance and system power flow. 

Once HIF occurs, parameters $\alpha_1, \cdots, \alpha_4$ are immediately altered, but as the circuit is not open,{\em{ the trajectory is still approximate elliptical with different parameters as $R_n,R_p$ vary}}.   

\begin{figure}[!ht] 
	\centering
	\includegraphics[width=0.24 \textwidth]{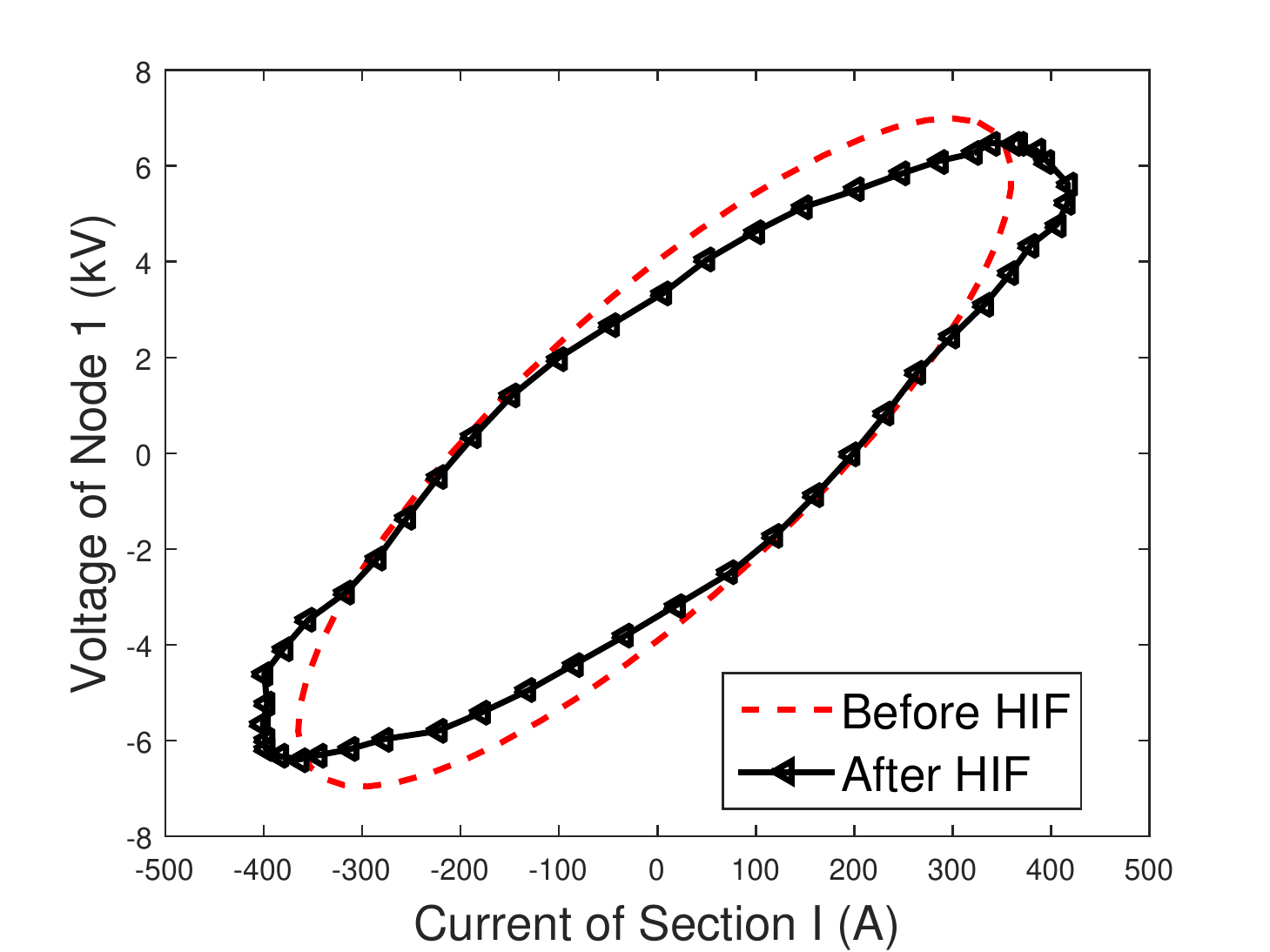}\hfill
	\includegraphics[ width=0.24 \textwidth]{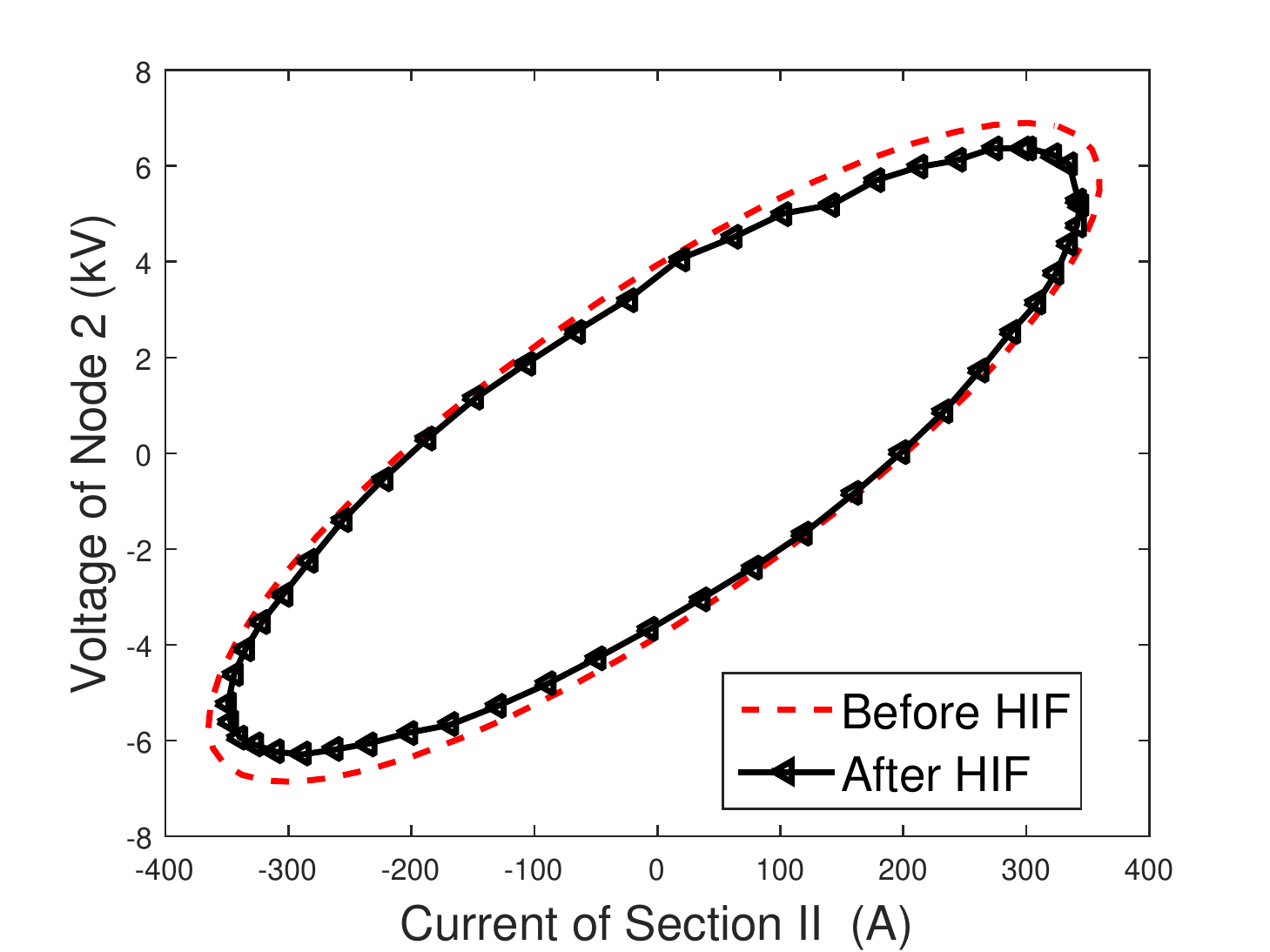} 
	\caption{ The trajectories of voltages and currents at node 1 and 2 in the four node test feeder system \cite{K91}, where the line  section I is between node 1 and 2 and section II is between node 2 and a transformer. Red curves are the voltages-current trajectories on normal conditions while the black ones are after HIF event }
	\label{fig:4ui} 
\end{figure}
\noindent \textbf{Four-node test feeder example:}  We illustrate the physical property of HIFs in  the four-node test feeder \cite{K91} simulated by PSCAD/EMTDC \cite{PSCAD}. When a HIF occurs near node 1, the trajectories of voltages and currents at node 1 and 2 are impacted.  

Fig.~\ref{fig:4ui} compares the trajectories before and after HIF event. It is clear that the black trajectory  deviates from the red one to formulate another approximate ellipse, and the deviation  is more serious when the node is closer to the HIF. 

As HIF's unique feature is the approximate elliptical trajectory of voltages and currents, varying from node to node, we present in the next section our detector that regulates the learning process in training by the elliptical trajectory without relying on sparsely available and expensive labels.  
\section{Physics-informed Convolutional Auto-encoder (PICAE)}\label{sec:method} 
\begin{figure}[!ht] 
	\centering
	\includegraphics[width=0.48 \textwidth]{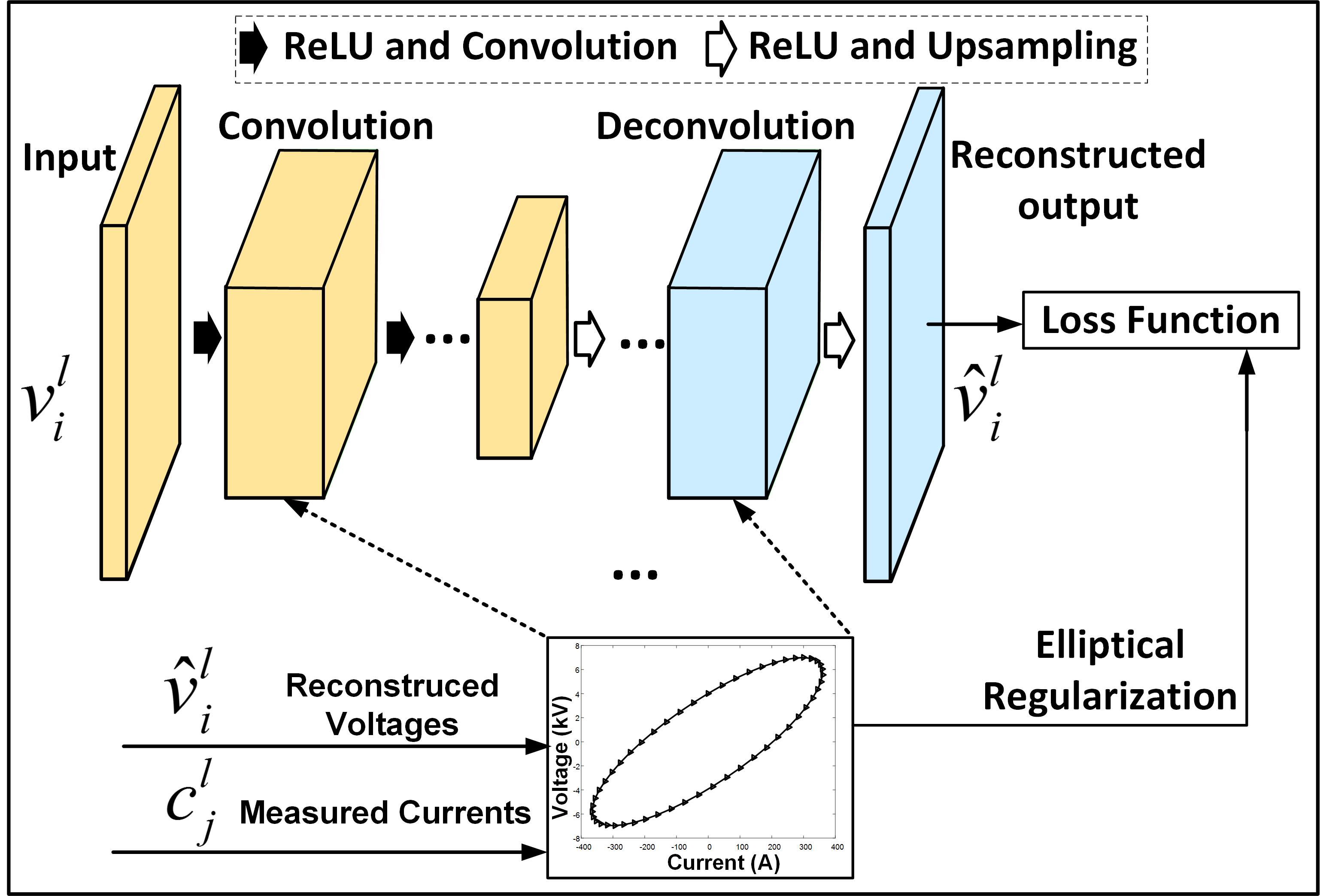} 
	\caption{ Physics-Informed Convolutional Auto-encoder} \label{PICAE} 
\end{figure} 
The configuration of our PICAE is shown in Fig.~\ref{PICAE}. Given time series of voltages a matrices $V_i \in R^{T \times N}, i=1, \cdots, m$ as inputs, where $m, T,N$ are the number of measured nodes, the length of the moving window,  and  the number of windows. According to the physical laws, the elliptical regularization structure in PICAE constrains the weights in the convolution and deconvolution layer. In training, the normal voltages  $V_i$ are reconstructed by the encoder, and then regulated by  the corresponding line currents to  obey the elliptical trajectory. Note that the elliptical regularization  is \emph{only employed during offline training}. In testing, only the measured voltages are needed to detect the occurrence of HIFs.

\subsection{The Encoder and Decoder of PICAE} \label{encoder}   
The encoder is a convolutional neural network with  decreasing size of  the latter layer than that of the previous layer.   The  $s$th convolutional layer down-samples the input $g^s$ with filters $W^s$ and bias matrices $B^s$ to reduce the dimensions and then goes through the nonlinear activation function of the Rectified Linear Units (ReLU) to enter into the next layer. 
\begin{align}\label{con}
	g^{s+1} = \max (0, g^{s} \circledast W^s + B^s), s=1, \cdots, S
\end{align}
where $\circledast$ denotes the convolution operation, and $g^1 = v^l_i \in R^{T }$, the $l$th column of $V_i$. 
The decoder has the symmetric structure with the encoder, which improves the reconstruction accuracy  \cite{GBC16}.  Here ``symmetric'' emphasizes the same sizes of the outputs of the deconvolution layer with that of the mirrored convolution layer. The  $h$th deconvolution layer up-samples the inputs $f^{h}$ with the filters $\bar{W}^h  $ and the bias $\bar{B}^h$ through deconvolutional and ReLU operations. 
\begin{align}\label{con}
	f^{h+1} & = \max (0, f^{h} \ast \bar{W}^h + \bar{B}^h), h=1, \cdots, S
\end{align}
where $\ast$ denotes the deconvolution operation. The final output $f^{S+1}$ is the reconstructed voltages $\hat{v}^l_i$. 

\subsection{Physical Regularization of PICAE} \label{physics} 
The regularization item acts as prior knowledge that direct the trained model to follow the latent physical rules mentioned in Section \ref{physical}, to enhance the robustness against noise and other abnormal events. Our regularization encodes the rotated elliptical trajectory of the nodal voltages against currents. Let time series $v_i$ be the voltage of the $i$th node in one window, and $ c_j \in R^T$ be the current on line connecting $i$ to a neighboring node $j \in \N(i)$. Let $Z_i = [v_i \odot v_i, v_i\odot c_j, c_j\odot c_j,v_i, c_j ] \in R^{T \times 5} $, where $\odot$ denotes the entry-wise product. Assuming normal conditions during the $T$ samples, the entries of voltages and currents measurements $v_i , c_j $ ideally follow an elliptical trajectory with five parameters $\beta = [a, b,c,d,e ]^T$, expressed as \cite{HF98}:
\begin{equation}\label{ellipse}
	Z_i \beta + {\bf f} = \bf 0 
\end{equation}
where ${\bf f}, \bf 0 \in R^T$ are an all one and all zero vectors, respectively. The \emph{ five unknown parameters in $\beta$} can be estimated by the following least square method, given sufficient number of voltages and currents measurements ($T \ge 5$):
\begin{align} \label{beta}
	\beta^{*} & = \arg \min_{\beta } \frac{1}{2} \lVert Z_i \beta + 
	{\bf f} \rVert_2^2= -(Z_i^T Z_i)^{-1} Z_i^T {\bf f}
\end{align} 
\textbf{Remark:} If no clean historical data-sets are present to compute $\beta^*$ through \eqref{beta}, we can approximate $\beta$ through power flow analysis. Specifically, as the equations of \eqref{parametric} and \eqref{ellipse} are equivalent, $\beta$ in \eqref{ellipse} can be estimated by the corresponding $V_0, C_0, \phi$ in \eqref{parametric} \cite{HF98}, obtained by power flow analysis on steady states \cite{LWZLK19}. 

\noindent \textbf{Training:} Given $N$ data samples $v_i^l,c_j^l,  l=1,\cdots, N$ of normal operation, the loss function of    PICAE for  node $i$ is:
\begin{align} \label{obj}
	\mathcal{L}(\Theta) = & \frac{1}{N } \Sigma_{l = 1}^N [\lVert v_i^l - \hat{v}^l_i(\Theta) \rVert_2^2 +  \lambda_r \lVert \hat{Z}_i \beta^{*} + 
	{\bf f} \rVert^2_2 ] 
\end{align}  

Here the first term $\lVert v_i^l - \hat{v}^l_i(\Theta) \rVert_2^2 $ denotes the mean square errors between the original and reconstructed voltages $\hat{v}^l_i(\Theta) $ with parameters $\Theta$.  The second item is the regularization, which uses the estimated $\beta^*$ to ensure that $\hat{v}_i^l $ follows the elliptical trajectory via $\hat{Z}^l_i = [\hat{v}^l_i \odot \hat{v}^l_i, \hat{v}^l_i\odot c_j^l, c_j^l\odot c_j^l,\hat{v}^l_i, c_j^l ]$. Considering the impact of topological changes in realistic power grids on $\beta$, $\lambda_r$ is set to be a relatively small value to allow some variations of the regularization term $\lVert \hat{Z}_i \beta^{*} +  {\bf f} \rVert^2_2 $, and the $\beta$ needs to be updated if the trajectory of voltages and currents changes significantly.   
The training also produces the average reconstructed error $\epsilon_i =\dfrac{1}{N} \Sigma_{l = 1}^N \lVert v_{i}^{l}- \hat{v}^{l}_{i} \rVert^2_2$ during normal conditions. The training steps are listed in Algorithm \ref{alg1}.
\begin{algorithm}[!ht] 
	\caption{Training of local  PICAE} \label{alg1}
	\begin{algorithmic}[1]
		\STATE Input: $N$ training datasets $v_i^l, c_j^l$, maximum iterations $k_{\max}$.
		\STATE Compute $\beta^{*} $ by \eqref{beta} with $v_i^l, c_j^l$; $k \gets 0$.
		\WHILE { $k < k_{\max}$ and early stop is not reached} 
		\STATE Optimize $\Theta $ of PICAE by minimizing $\mathcal{L}( \Theta) $ in \eqref{obj}. 
		\ENDWHILE
		\STATE Output: trained  PICAE, $\epsilon_i =\dfrac{1}{N} \Sigma_{l = 1}^N \lVert v_{i}^{l}- \hat{v}^{l}_{i} \rVert_2$ on normal conditions. 
	\end{algorithmic} 
\end{algorithm} 
In \emph{testing}, we use the trained  PICAE on online voltage $v_i^{l'}$ to reconstruct voltages, and determine the confidence score $\gamma_i= {\varepsilon_i}/{\epsilon_i}$, the relative error compared to testing, where $\varepsilon_i = \lVert v_i^{l'}- \hat{v}^{l'}_{i} \rVert^2_2$ is the mean square reconstructed error of the testing data $v_i^{l'}$. We compare   $\gamma_i$  with two predefined thresholds $\xi_1, \xi_2$  to distinguish HIFs from other events. The threshold  $\xi_1 = \frac{ \max_l \lVert v_{i}^{l}- \hat{v}^{l}_{i} \rVert^2_2}{\epsilon_i}$ to discern the normal and abnormal events based on the results of  training in Algorithm \ref{alg1}. Then if  $\gamma_i $ of the testing data $v_i^{l'}$ is lower than $\xi_1$, the algorithm treats the testing data  as normal  since the PICAE can well represent normal voltages with a small  reconstruction error; another threshold $\xi_2$ is defined with the maximum confidence score computed  by validation data samples of a few HIF events.  As HIFs follow the elliptical trajectory, explained in Section \ref{sec:model}, the reconstruction errors of HIFs are smaller than those of events, such as capacitor switching, disobeying the elliptical trajectory, which is further explained in Section \ref{regu} in details.  Thus the primary function  of $\xi_2$ is to distinguish HIFs from other abnormal events.   The detailed steps are listed in Algorithm \ref{alg2}.  
\begin{algorithm}[!ht] 
	\caption{HIF detection through Local  PICAE} \label{alg2}
	\begin{algorithmic}[1]
		\STATE Input: Online testing dataset in moving windows $v_i^{l'}, l'=1,\cdots, N'$, averaged reconstruction error $ \epsilon_i$ for normal voltages of node $i$, two thresholds $ \xi_1, \xi_2$. \\ 
		\STATE Input $v_i^{l'}$ into trained PICAE to reconstruct $\hat{v}^{l'}_{i}$. 
		\STATE $\varepsilon_i \gets \lVert v_i^{l'}- \hat{v}^{l'}_{i} \rVert^2_2$. Confidence score $\gamma_i \gets {\varepsilon_i}/{\epsilon_i} $
		\IF{$ \gamma_i < \xi_1 $ }
		\STATE Output: normal conditions
		\ELSIF{$\gamma_i > \xi_2$}
		\STATE Output: Other abnormal events are detected
		\ELSE 
		\STATE Output: HIF events are detected
		\ENDIF
	\end{algorithmic} 
\end{algorithm} 

\section{Centralized HIF Detection Framework  for Partially Observed Systems } \label{sec:detect}
\vspace{-3mm}
\begin{figure}[!ht] 
	\centering
	\includegraphics[width=0.4 \textwidth]{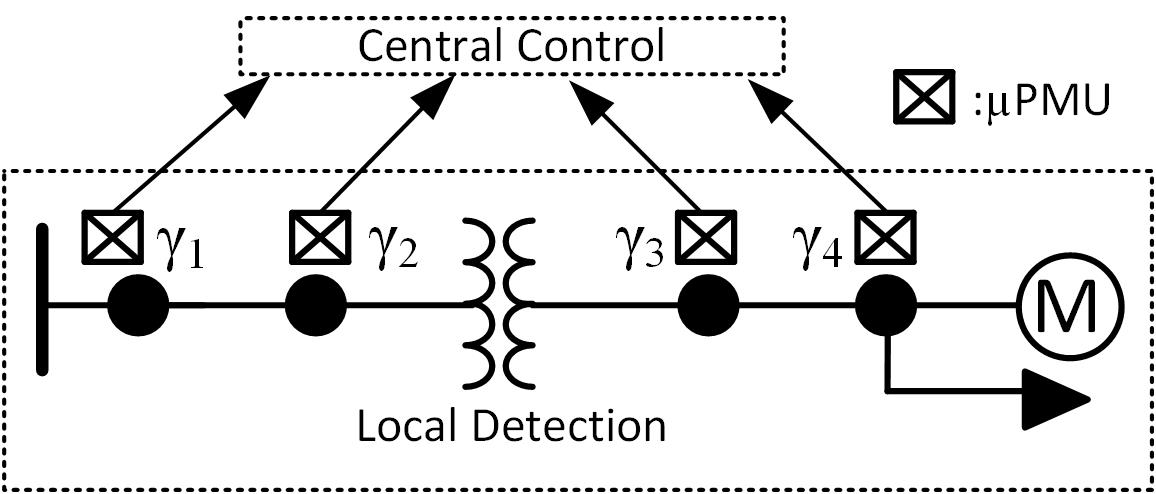} 
	\caption{ The configuration of the proposed detection framework. $\gamma_i $ is the confidence score of the $i$th measured node for HIF detection} \label{fig:conf}
\end{figure}
While Algorithm \ref{alg2} is implemented for each observed node independently, we  design a system-wise detection framework in Fig.~\ref{fig:conf} combining all the local detectors for the partially observed systems. The computed $\gamma_i$ at each observed node can be communicated to a central detector (Distribution system operator), which decides HIF occurrence using $\max \gamma_i$. Note that we avoid high communication overhead by not relying on the entire voltage sequence to the central detector. The high $\gamma_i$ scores can also provide auxiliary information about the possible location of the HIF since we observe that the nearby node voltages reveal a relatively high confidence score. 

\section{Numerical Experiments}\label{sec:simu}
\begin{figure}[!ht] 
	\centering
	\includegraphics[width=0.48\textwidth]{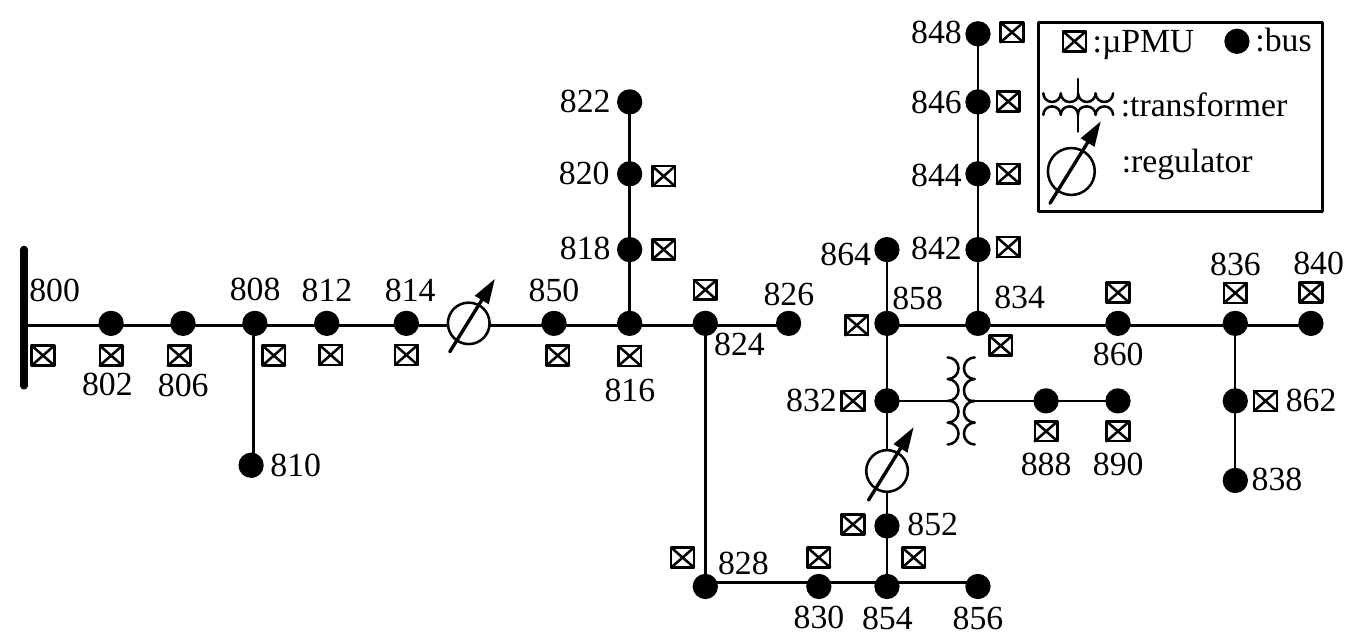} 
	\caption{ 34 node testing feeder \cite{MS07} } \label{34node} %
\end{figure}
\begin{table} 
	\centering
	\caption{The variation range of parameters of HIF model }
	\label{tab:para}
	\begin{tabular}{c |c |c| c c }
		\hline \hline $R_p (\Omega)$ & $R_n (\Omega)$ & $ V_p (kV)$ & $V_n (kV)$ \\ 
		\hline 600 $\sim$ 1400 & 600 $\sim$ 1400 & 5 $\sim$ 6 & 7 $\sim$ 8 \\
		\hline
	\end{tabular}
\end{table}
We validate our approaches in the IEEE 34-node in Fig.~\ref{34node} with a voltage level of 24.9 kV test feeder \cite{MS07} modeled by PSCAD/EMTDC\cite{PSCAD}. The parameters $R_p, R_n, V_p, V_n$ of HIF models  vary  in the ranges of Table \ref{tab:para} randomly at every 1K Hz  \cite{CD18, GB12}.  We record wave-forms of node voltages and line currents with 512 samples per cycle, and the interval between any two consecutive windows is around four million-second (ms) or $\tau = 128$ samples. Training datasets are composed of $N = 325$ windows of node voltages and  line currents. Total 286 testing events in various situations include: 100 HIF events occurring on different branches with varying resistance and DC voltages; 42 different loads switching near the node 890 at various time instants, 54 capacitor switching near the node 844 with the reactive powers in the range of 0.5 to 5 MVA; the remaining 90 normal events with varying initial conditions. We also generate another 10\% of testing HIFs events with different random parameters as the validation data for the model   selection described in Section III-B. We apply the range normalization to augment the data-sets \cite{ZMbook14}. The designed PICAE has the symmetrical two convolution-layer structure with the filter $W^s$ size of 5.  The number of filters of the two layers change from 32 to 1 to generate hidden variables in a low-dimension subspace. We train the PICAE using the Adam optimizer \cite{KBJ14} with a learning rate of 0.0001 and batch of size 12. The maximum iteration $k_{\max} = 1500$ and $\lambda_r = 200$ in \eqref{obj}. Note that we present our major results here, but more extensive explanations and experimental results are in the supplement materials \cite{LD20}.
\subsection{Performance Metrics} 
We evaluate the detection performance with three criteria: \textbf{Precision, Recall and F1 score }\cite{MK12}.   A high ``precision'' demonstrates that the detector has a low mistake rate of identifying non-HIF as HIF events. A large ``recall'' value means that the detector has a strong capability to recognize HIF events from others. ``F1 score''  is a weighted average of precision and recall, and comprehensively evaluates the capability of the detector. 
\subsection{Detection Performance with Partial Measurements} 
To investigate the   detection performance   for the distribution system without full observability, we show the detection performance when only 24\% to 6\% (or 8 to 2) nodes are measured. We compare the detection performance when the placement of the measured nodes are ``random'' (the averaged performance after 100 times of uniformly random selection) and ``selected'' (determined by the algorithm in \cite{LD20} ).   The ``recall'' degrades for the low measured ratio, because some abnormal events, such as capacitor switching,  are far away from the measured nodes that their signatures are not fully captured,    but the proper placement of $\mu$PMU improves the performance by 1\%$\sim$15\%.  Table \ref{tab:placement} reveals that the detection performance can be more than 95\% when more than 24\%  nodes are measured.
\begin{table}[!h]
	\centering
	\caption{Detection Performance with different $\mu$PMU placement algorithms when system is partial observed with $\xi_1 = 2, \xi_2 = 350$} 
	\label{tab:placement}
	\begin{tabular}{c |c |c| c }
		\hline \hline Measured Ratio & Precision & Recall & F1 score\\ 
		\hline 24\% (Selected) &  \bf 100.0\% &   \bf  	100.0\% &  	 \bf 100.0\%	 \\ 
		\hline 24\% (Random) &  	100.0\% & 98.1\% &	99.0\%\\ 
		\hline 12\%   (Selected) &  \bf 	100.0\% &   \bf  98.0\% &  \bf  	99.0\% \\ 
		\hline 12\% (Random) & 	100.0\% & 87.7\% & 	93.1\% \\ 
		\hline 6\%  (Selected) &  \bf  100.0\%&   \bf 92.0\%  &  \bf   	95.8\% \\
		\hline 6\% (Random) & 	100.0\% & 68.5\% & 	80.3\% \\ 
		\hline 
	\end{tabular}
\end{table}   
\subsection{The Effectiveness of the Regularization } \label{regu}
\vspace{-3mm}
\begin{figure}[!ht] 
	\centering 
	\includegraphics[width=0.23\textwidth]{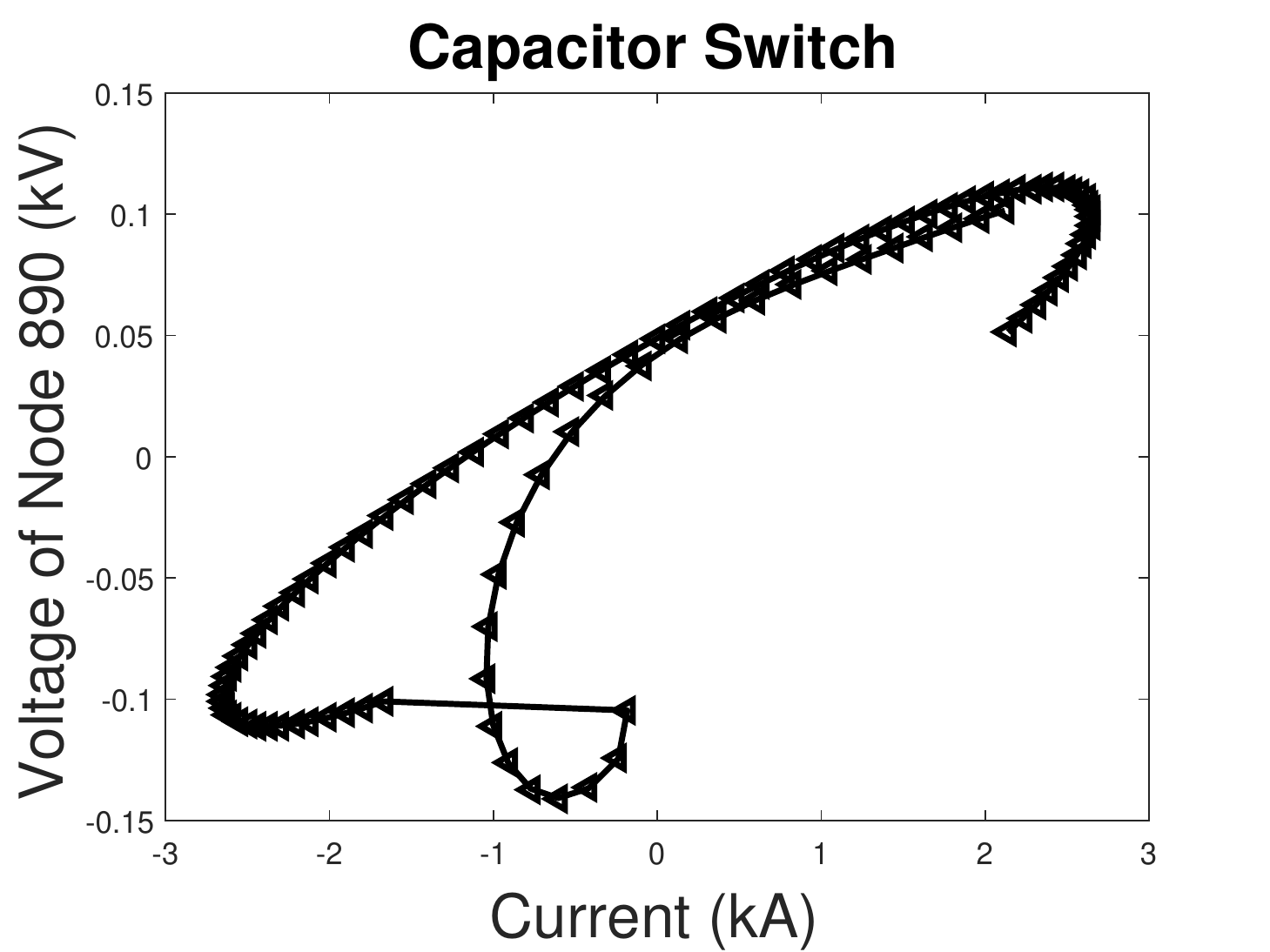} 
	\includegraphics[width=0.23\textwidth]{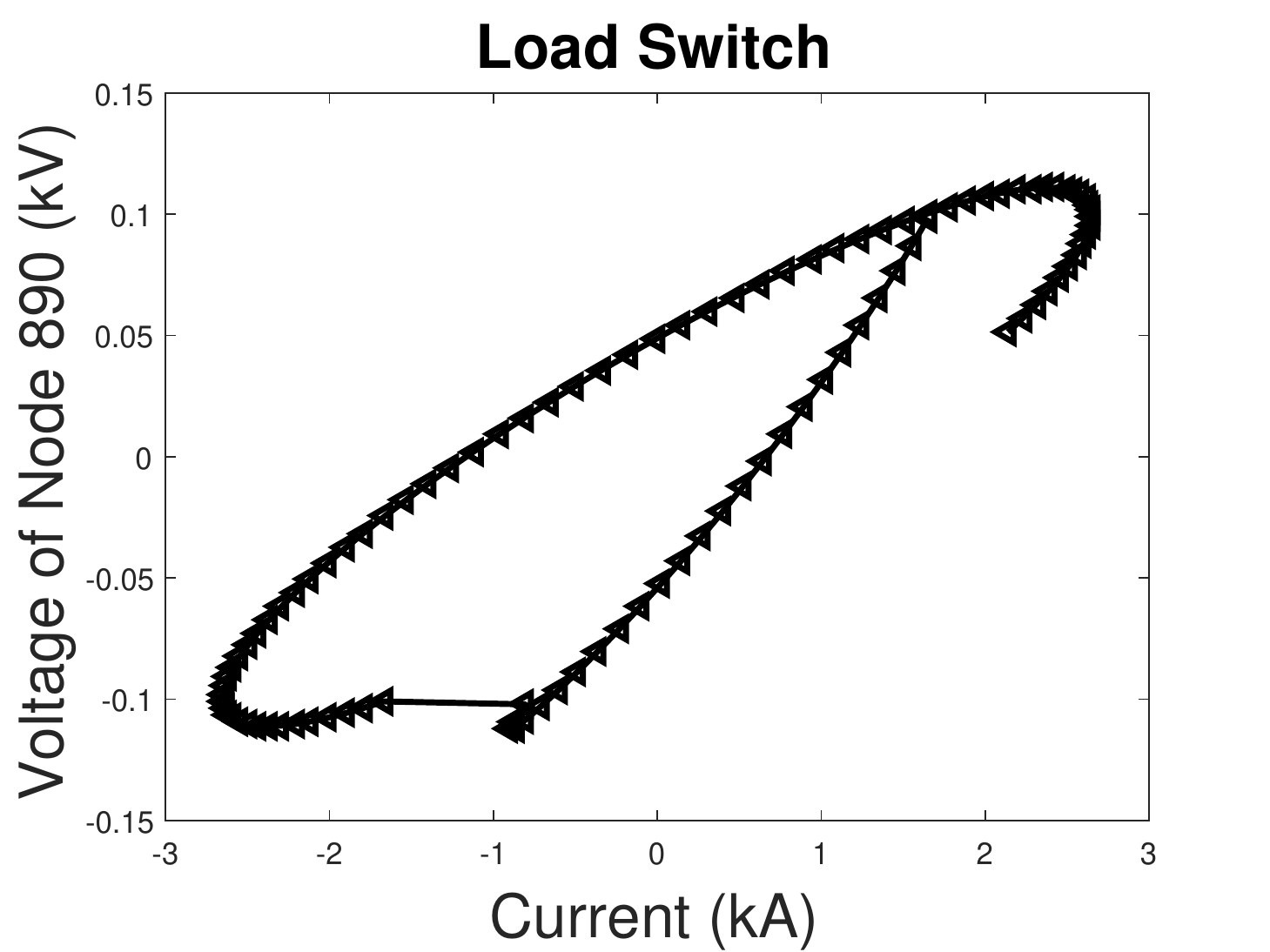} 
	\caption{ The characteristics curves of voltages currents in one cycle at the node 890 when a capacitor bank switch or a load switch occurs respectively } \label{fig:traj} 
\end{figure} 
\begin{figure}[!ht] 
	\centering 
	\includegraphics[width=0.5\textwidth]{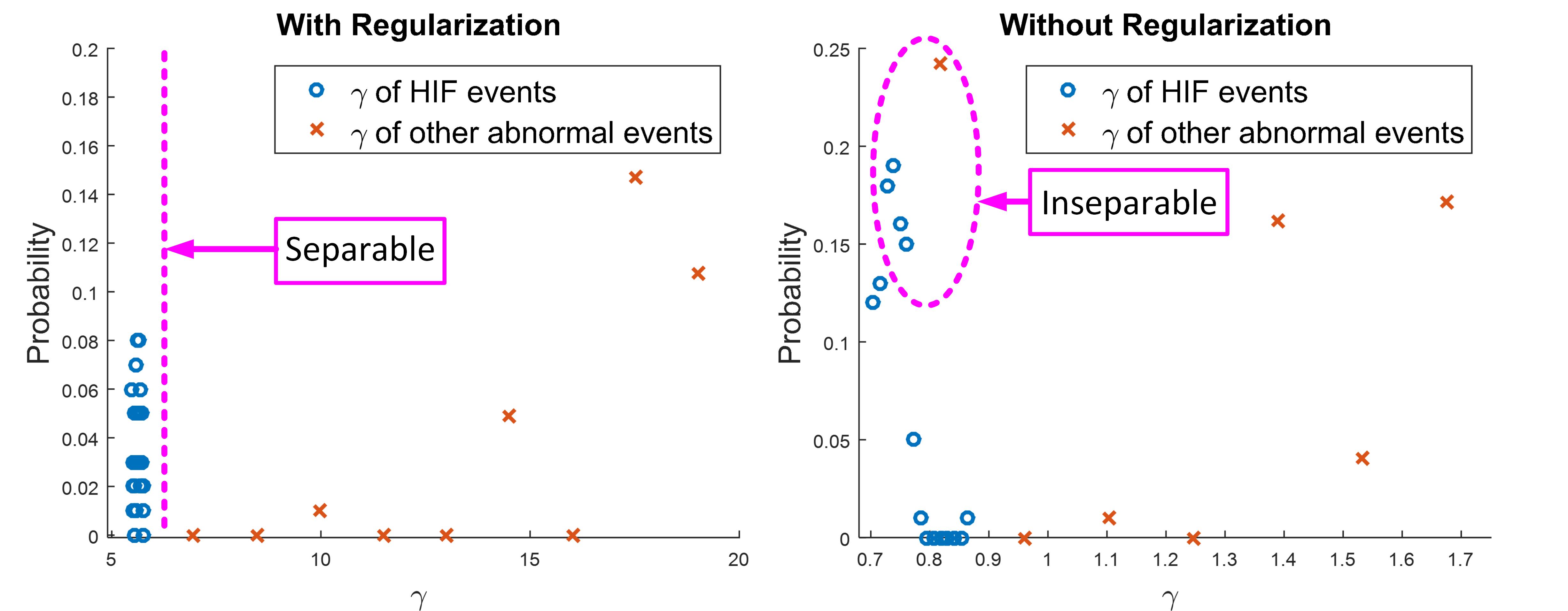} 
	\caption{Probability of $\gamma$ for different noisy abnormal events when SNR is 50dB detected by PICAE with (left) and without (right) regularization item } \label{fig:prob} 
\end{figure} 
Fig.~\ref{fig:traj} displays the trajectories of voltages and currents in one cycle after capacitor switching and load switching occur respectively. It is evident that these trajectories deviate from the original ellipse dramatically. On the contrary, Fig.~\ref{fig:4ui} indicates that the trajectories for HIFs still follow certain ellipses. As a result, the reconstruction errors of PICAE for the capacitor switching and loads switching are significant compared with that of HIFs. Hence, the reconstruction errors themselves distinguish HIFs from other abnormal events due to the \emph{elliptical regularization item}. 

We discover this separability of PICAE becomes even more evident in noisy situations. According to the practical noise level of PMUs \cite{MB2016}, we corrupt the training and testing datasets by Gaussian noise of signal-noise-ratio (SNR) ranging from 30 dB to 90 dB and train the PICAEs with ($\lambda_r \neq 0$) and without ($\lambda_r = 0$) the regularization item. Fig.~\ref{fig:prob} statistically depicts the probabilistic distribution of $\gamma$'s, which generally reflect the variations of reconstruction errors, 
of various testing events in noisy situations. 
The $\gamma$'s of the HIFs become separable from those of the other abnormal events when using the PICAE with the elliptical regularization. On the contrary, the HIFs and non-HIFs are not separable if the PICAE is trained without the regularization. 
\subsection{Comparison with Existing Works}
\begin{table}[!ht] 
	\centering
	\caption{Detection F1 score of the PICAE for node 832 when SNRs are from 30dB to 90dB } \label{tab:F1}
	\begin{tabular}{c|c |c |c|c }
		\hline \hline SNR (dB) & PICAE & AE & PCA & ER \\ 
		\hline 
		30dB &  \textbf{92.9\% } 	 & 81.5\% & 43.2\% & 39.5\% \\
		\hline 
		50dB & 	\textbf{97.1\%} & 	81.3\% & 72.2\% & 62.9\% \\
		\hline 
		70dB & 	 \textbf{97.6\%} & 	83.0\% & 76.1\% & 
		64.7\% \\
		\hline 
		90dB & 	\textbf{100.0\%} & 	83.3\% & 	76.6\% & 
		64.7\% \\
		\hline 
	\end{tabular} 
\end{table} 
We compare the detection performance of the local PICAE with three existing unsupervised methods: auto-encoder (AE), principle component analysis (PCA), and Ellipse regression (ER) \cite{GBC16, MK12, RC14}. The structure of AE is similar to PICAE but without the physical regularization. We implement the PCA by the truncated singular value decomposition (SVD), and the number of principle components is selected by $r^* = \arg \min_{r} \dfrac{\Sigma_{n = 1}^r \sigma_n}{\Sigma_{n = 1}^T \sigma_n} \geq \tau$, where $\sigma_n$'s are the decreasing singular values of voltages $V_i$, and $\tau = 0.99$. ER represents the training data using the  elliptical equation \eqref{ellipse}, through a linear regression method \cite{RC14}. The performance of these three methods for normal and abnormal events are employed in the same way of Algorithm \ref{alg2} to detect HIFs. 

We summarize the F1 score of these four methods when SNR changes from 30 dB to 90 dB in Table \ref{tab:F1}.   PICAE is more robust to noise than others, achieving up to 17\% higher F1 scores above all. The improvement profits from two attributes of PICAE. First, the convolutional autoencoder reconstructs normal events with high accuracy. Second, the physical regularization term enables a more considerable distinction between HIFs and other non-HIFs even in noisy situations. Note that when the SNR as low as 30 dB, we increase $\lambda_r = 440$ to improve the detection performance.  
\section{Conclusions} \label{sec:con}
HIF, potentially causing wildfires in the western U.S., is a significant concern in the industry.  Existing data-driven algorithms can detect HIFs with high accuracy when a sufficient number of labeled datasets are provided. Rather than relying on the expensive labeled datasets, our PICAE exploits the unique voltage-current characteristic curves of HIFs as regularization in training. The regularization improves the capability of PICAE to separate  HIFs from other events, even in highly noisy situations. Furthermore, a low-communication system-wide detection framework is proposed   to improve detection accuracy and reliability, especially when systems have low observability. PICAE demonstrates superior performances in different noisy scenarios than existing works. An interesting avenue for future work is to unify the location and detection algorithms to enable follow-up control actions.

\bibliographystyle{IEEEtran} 
\bibliography{IEEEabrv,ref,WentingPub}   

\begin{thebibliography}{10}
\providecommand{\url}[1]{#1}
\csname url@samestyle\endcsname
\providecommand{\newblock}{\relax}
\providecommand{\bibinfo}[2]{#2}
\providecommand{\BIBentrySTDinterwordspacing}{\spaceskip=0pt\relax}
\providecommand{\BIBentryALTinterwordstretchfactor}{4}
\providecommand{\BIBentryALTinterwordspacing}{\spaceskip=\fontdimen2\font plus
\BIBentryALTinterwordstretchfactor\fontdimen3\font minus
  \fontdimen4\font\relax}
\providecommand{\BIBforeignlanguage}[2]{{%
\expandafter\ifx\csname l@#1\endcsname\relax
\typeout{** WARNING: IEEEtran.bst: No hyphenation pattern has been}%
\typeout{** loaded for the language `#1'. Using the pattern for}%
\typeout{** the default language instead.}%
\else
\language=\csname l@#1\endcsname
\fi
#2}}
\providecommand{\BIBdecl}{\relax}
\BIBdecl

\bibitem{GB12}
S.~Gautam and S.~M. Brahma, ``Detection of high impedance fault in power
  distribution systems using mathematical morphology,'' \emph{{IEEE} Trans.
  Power Syst.}, vol.~28, no.~2, pp. 1226--1234, 2012.

\bibitem{MRB96}
A.~Mamishev, B.~D. Russell, and C.~L. Benner, ``Analysis of high impedance
  faults using fractal techniques,'' \emph{{IEEE} Trans. Power Syst.}, vol.~11,
  no.~1, pp. 435--440, 1996.

\bibitem{GMGS14}
A.~Ghaderi, H.~A. Mohammadpour, H.~L. Ginn, and Y.-J. Shin, ``High-impedance
  fault detection in the distribution network using the time-frequency-based
  algorithm,'' \emph{{IEEE} Trans. Power Del.}, vol.~30, no.~3, pp. 1260--1268,
  2014.

\bibitem{AWTJ06}
M.~Adamiak, C.~Wester, M.~Thakur, and C.~Jensen, ``High impedance fault
  detection on distribution feeders,'' \emph{GE Industrial solutions}, 2006.

\bibitem{WGD16}
B.~Wang, J.~Geng, and X.~Dong, ``High-impedance fault detection based on
  nonlinear voltage--current characteristic profile identification,''
  \emph{{IEEE} Trans. Smart Grid}, vol.~9, no.~4, pp. 3783--3791, 2016.

\bibitem{CD18}
S.~Chakraborty and S.~Das, ``Application of smart meters in high impedance
  fault detection on distribution systems,'' \emph{{IEEE} Trans. Smart Grid},
  vol.~10, no.~3, pp. 3465--3473, 2018.

\bibitem{LBHG15}
Y.~LeCun, Y.~Bengio, and G.~Hinton, ``Deep learning,'' \emph{Nature}, vol. 521,
  no. 7553, pp. 436--444, May 2015.

\bibitem{PGHY16}
Y.~Pu, Z.~Gan, R.~Henao, X.~Yuan, C.~Li, A.~Stevens, and L.~Carin,
  ``Variational autoencoder for deep learning of images, labels and captions,''
  in \emph{Proc. Adv. Neural Inf. Process. Syst.}, 2016, pp. 2352--2360.

\bibitem{MSY16}
X.~Mao, C.~Shen, and Y.-B. Yang, ``Image restoration using very deep
  convolutional encoder-decoder networks with symmetric skip connections,'' in
  \emph{Proc. Adv. Neural Inf. Process. Syst.}, 2016, pp. 2802--2810.

\bibitem{ZS20}
A.~S. Zamzam and N.~D. Sidiropoulos, ``Physics-aware neural networks for
  distribution system state estimation,'' \emph{{IEEE} Trans. Power Syst.},
  2020.

\bibitem{PDC18}
S.~Park, D.~Deka, and M.~Chcrtkov, ``Exact topology and parameter estimation in
  distribution grids with minimal observability,'' in \emph{2018 Power Systems
  Computation Conference (PSCC)}.\hskip 1em plus 0.5em minus 0.4em\relax IEEE,
  2018, pp. 1--6.

\bibitem{TDDMCS20}
S.~Talukdar, D.~Deka, H.~Doddi, D.~Materassi, M.~Chertkov, and M.~V. Salapaka,
  ``Physics informed topology learning in networks of linear dynamical
  systems,'' \emph{Automatica}, vol. 112, p. 108705, 2020.

\bibitem{LWC18}
W.~Li, M.~Wang, and J.~H. Chow, ``Real-time event identification through
  low-dimensional subspace characterization of high-dimensional synchrophasor
  data,'' \emph{{IEEE} Trans. Power Syst.}, vol.~33, no.~5, pp. 4937--4947,
  Jan. 2018.

\bibitem{LDCW19}
W.~{Li}, D.~{Deka}, M.~{Chertkov}, and M.~{Wang}, ``Real-time faulted line
  localization and pmu placement in power systems through convolutional neural
  networks,'' \emph{{IEEE} Trans. Power Syst.}, vol.~34, no.~6, pp. 4640--4651,
  Nov. 2019.

\bibitem{RPK17}
M.~Raissi, P.~Perdikaris, and G.~E. Karniadakis, ``Physics informed deep
  learning (part i): Data-driven solutions of nonlinear partial differential
  equations,'' \emph{arXiv preprint arXiv:1711.10561}, 2017.

\bibitem{MS07}
N.~Mwakabuta and A.~Sekar, ``Comparative study of the ieee 34 node test feeder
  under practical simplifications,'' in \emph{2007 39th North American Power
  Symposium}.\hskip 1em plus 0.5em minus 0.4em\relax IEEE, 2007, pp. 484--491.

\bibitem{PSCAD}
\emph{PSCAD/EMTDC manual version}, 4th~ed., Manitoba H. V. D. C. Research
  centre Inc., 211 Commerce Drive Winnipeg, Manitoba Canada R3P 1A3, 2018.

\bibitem{SS90}
S.~Shiller, ``High impedance fault arcing on sandy soil in 15kv distribution
  feeders: contributions to the evaluation of the low frequency spectrum,''
  \emph{{IEEE} Trans. Power Del.}, vol.~5, no.~2, 1990.

\bibitem{TPV18}
J.~J. Theron, A.~Pal, and A.~Varghese, ``Tutorial on high impedance fault
  detection,'' in \emph{2018 71st Annual Conference for Protective Relay
  Engineers (CPRE)}.\hskip 1em plus 0.5em minus 0.4em\relax IEEE, 2018, pp.
  1--23.

\bibitem{K91}
W.~H. Kersting, ``Radial distribution test feeders,'' \emph{{IEEE} Trans. Power
  Syst.}, vol.~6, no.~3, pp. 975--985, 1991.

\bibitem{GBC16}
I.~Goodfellow, Y.~Bengio, and A.~Courville, \emph{Deep Learning}.\hskip 1em
  plus 0.5em minus 0.4em\relax Cambridge, MA, USA: MIT Press, 2016.

\bibitem{HF98}
R.~Hal{\i}r and J.~Flusser, ``Numerically stable direct least squares fitting
  of ellipses,'' in \emph{Proc. 6th International Conference in Central Europe
  on Computer Graphics and Visualization. WSCG}, vol.~98.\hskip 1em plus 0.5em
  minus 0.4em\relax Citeseer, 1998, pp. 125--132.

\bibitem{LWZLK19}
Y.~Liu, Y.~Wang, N.~Zhang, D.~Lu, and C.~Kang, ``A data-driven approach to
  linearize power flow equations considering measurement noise,'' \emph{{IEEE}
  Trans. Smart Grid}, vol.~11, no.~3, pp. 2576--2587, 2019.

\bibitem{ZMbook14}
M.~J. Zaki and W.~Meira, \emph{Data mining and analysis: fundamental concepts
  and algorithms}.\hskip 1em plus 0.5em minus 0.4em\relax Cambridge University
  Press, 2014.

\bibitem{KBJ14}
D.~P. Kingma and J.~L. Ba, ``Adam: Amethod for stochastic optimization,'' in
  \emph{Proc. 3rd Int. Conf. Learn. Representations}, 2014, pp. 1--15.

\bibitem{LD20}
W.~Li and D.~Deka, ``Physics regulated neural network for high impedance faults
  detection,'' \emph{arXiv preprint arXiv:2008.02364}, 2020.

\bibitem{MK12}
K.~P. Murphy, \emph{Machine Learning: A Probabilistic Perspective}.\hskip 1em
  plus 0.5em minus 0.4em\relax Cambridge, MA, USA: MIT Press, 2012.

\bibitem{MB2016}
S.~B. S. J.~R. Michael~Brown, Milan~Biswal and H.~Cao, ``Characterizing and
  quantifying noise in {PMU} data,'' in \emph{Proc. IEEE Power and Energy
  Society General Meeting}, 2016, pp. 1--5.

\bibitem{RC14}
C.~Robert, \emph{Machine learning, a probabilistic perspective}.\hskip 1em plus
  0.5em minus 0.4em\relax United Kingdom: Taylor \& Francis, 2014.

\bibitem{DV11}
D.~Deka and S.~Vishwanath, ``{PMU} placement and error control using belief
  propagation,'' in \emph{Proc. IEEE Int. Conf. Smart Grid Communications},
  2011, pp. 552--557.

\bibitem{NK13}
K.~G. Nagananda, ``Electrical structure-based {PMU} placement in electric power
  systems,'' \emph{arXiv preprint arXiv:1309.1300}, 2013.

\bibitem{SM2014}
M.~Soltanolkotabi \emph{et~al.}, ``Robust subspace clustering,'' \emph{Ann.
  Stat.}, vol.~42, no.~2, pp. 669--699, April 2014.

\end{thebibliography}

\appendix

In the realistic setting where $\mu$PMUs and the corresponding PICAEs are sparsely placed in the distribution grid, the centralized HIF detector's performance depends on the placement of the $ \mu$PMUs. We now discuss a $\mu$PMU placement algorithm to maximize the detection performance using a limited number of $K$ observed nodes. 
\subsection{$\mu$PMU Placement Algorithm } \label{sec:pmu}
The placement of $\mu$PMU is crucial because the signatures of HIFs are local and only revealed by nearby $\mu$PMUs. Conventional PMU or $\mu$PMU placement algorithms determine PMU placement by solving a set cover problem \cite{DV11},\cite{NK13}, that ensures that each bus is within one-hop of a PMU, or at least one terminal bus of a line has a PMU. In settings where the number of PMUs is too small to ensure complete observability, we present an alternate placement approach that maximizes the recorded PMU data diversity to improve detection.

The intuition comes from the empirical observation that grid segments/edges have distinctive voltages-curves at different parts of the network.   By collecting measurements from nodes with different voltage dynamics, we are able to model the diversity of features. We measure the distinction of the voltages $v_i$ and $v_j$ by the subspace angle $\delta_{i,j}$ \cite{SM2014},  
\begin{equation}\label{simi}
\delta_{i,j} = \begin{cases}
\arccos( \frac{ \cos(v_i, v_j)}{\lVert v_i \rVert_2 \lVert v_j \rVert_2}) & \quad \text{if } (i,j) \in \E \\
0 & \quad \text{else } 
 \end{cases} 
\end{equation}
where we only compare the dissimilarity of nodes $i$ and $j$ if $(i, j) \in \E$. With the measured distinction $\delta_{i, j}$, we determine a set $\S$ of at most $K$ non-adjacent $\mu$PMU locations that maximizes the total dissimilarity $\Sigma_{i \in \S, j \in \N(i) } \delta_{i,j}$. 
Algorithm \ref{alg} provides a greedy approach to determine locations to  maximize the total dissimilarity. The performance improvements due to our placement strategy is described with other numerical experiments in the next section. 
\begin{algorithm}[!ht] 
\caption{ $\mu$PMU Placement } \label{alg}
\begin{algorithmic}[1]
\STATE Input: $K, \delta_{i,j}, i, j = 1, \cdots, m $ 
\STATE $\S \gets \emptyset$, $\Delta_i = \Sigma_{ j \in \N(i) } \delta_{i,j} $. 
\WHILE { $|S| < K$ and $\Delta_i$ is not a all-zero vector}
\STATE $\S \gets \S \cup i^*, \Delta_{j} \gets 0, \forall j \in \N(i^*)$, where $i^* = \arg \max_{i}\Delta_{i}$ 
\ENDWHILE
\STATE Output: $\S$
\end{algorithmic} 
\end{algorithm}

\subsection{Robustness to Low Sampling rates} \label{sect:pdf} 
\begin{table}[!ht]
\centering
\caption{Detection Performance of local PICAE at node 832 for Different Low Sampling Rate}
\label{tab:sample}
\begin{tabular}{c|c |c |c| c |c}
\hline  
\hline $f$ (kHz) & 15.36 & 7.68 & 3.84 &	1.92 & 0.96 \\ 
\hline $T$ & 256 & 128 & 64 & 32 & 16 \\
\hline Precision (\%)  & 100.0\%	 & 	100.0\%	 & 	100.0\%	 & 	95.2\%	 & 	94.3\%	 \\ 
\hline Recall  (\%)  & 100.0\%	 & 100.0\%	 & 	100.0\%	 & 		100.0\%		 & 	100.0\%			\\  
\hline F1 Score  (\%)  & 100.0\%	 & 	100.0\%	 & 	100.0\%	 & 	97.6\% 	 & 	97.1\%  \\ 
\hline 
\end{tabular}
\end{table} 
 We downsample the datasets and demonstrate the robustness of PICAE to low sampling rates in Table~\ref{tab:sample}, which is one of the concerns in the industry. When $T$, the number of samples per cycle, changes from 256 to 16, F1 score of the PICAE is higher than 90\%, indicating the same PICAE tolerates lower sampling rates without obvious reduction of accuracy. Moreover, the structure of PICAE adapts to inputs of various sampling rate and does not require redesigning of the filters and bias matrices.

\end{document}